\documentstyle[pre,aps,eqsecnum,epsf,graphicx]{revtex}
\begin{document}

\draft

\title{Momentum dependence of symmetry potential in asymmetric nuclear matter
for transport model calculations}

\author{C. B. Das$^1$, S. Das Gupta$^1$, C. Gale$^1$ and Bao-An Li$^2$}

\address{
$^1$Physics Department, McGill University, Montr{\'e}al, Canada H3A 2T8}
\address{$^2$Department of Chemistry and Physics, Arkansas State
University, Arkansas 72467-0419, USA}

\date{\today}

\maketitle

\begin{abstract}
For transport model simulations of collisions between two nuclei which
have $N/Z$ significantly different from unity one needs a one-body
potential which is both isospin and momentum dependent.  This
work provides sets of such potentials.

\end{abstract}


\section{Introduction}
Momentum dependent mean-fields for transport model calculations of heavy ion
collisions have been in usage for many years now 
\cite{Gale1,Prakash,Welke,Gale2,Pan,Zhang}.  So far the attention has been
for a momentum dependent potential which does not distinguish between
neutrons and protons. This is adequate for systems which have $N\approx Z$
where $N$ is the number of neutrons and $Z$ the number of protons.
One main focus of nuclear physics research today is to explore both structures of 
exotic nuclei in regions far-off the stability line and novel 
properties of neutron-rich nuclear matter. The latter can be investigated by using 
collisions induced by neutron-rich nuclei at intermediate to high energies. 
To interpret critically data from these collisions and
to extract accurately properties of neutron-rich nuclear matter, advanced transport model 
calculations are necessary. In asymmetric nuclear matter, the one body potential seen by a proton 
is different from that seen by a neutron. This has been implemented in 
BUU (Boltzmann-Uehling-Uhlenbeck) calculations but with a simplification that the 
potentials are taken to be momentum independent \cite{Li}. The present work aims to 
correct this deficiency. That the momentum dependence will be different for neutrons and
protons is of course well-known and has been the subject of quite
sophisticated many body calculations, see, e.g, ref.\ \cite{Bombaci} for a recent review.  
We do not aim to add anything fundamental in this regard.  Our objective is to
obtain a parametrized version which displays the main characteristics
of momentum dependence in asymmetric matter and is still usable in
practical BUU calculations. Major advances in this direction were already made: see articles
by Bombaci \cite{Bombaci} and Prakash et al.\cite{Prakash2}.
We add to this.  We will not only extend the simplest
momentum dependent potential \cite {Gale1} to include isospin but
also extend the improved treatment \cite{Prakash,Welke,Gale2} 
subsequently introduced to include isospin.  Thus this is an extension
of the work reported in \cite {Welke,Gale2}.

\section{A momentum dependent potential from a phenomenological
interaction}
An effective momentum dependent potential can be deduced from phenomenological
interactions.  We take the Gogny interaction \cite{Gogny} to obtain an
idea of the momentum dependence.  There are many reasons for this choice.
It has been used in detailed fits for spectra in finite nuclei.
It gives accepted values for binding energy, saturation density,
compressibility and symmetry energy in nuclear matter.
It has been verified already \cite{Prakash} that
the interaction produces a reasonable parameterisation for the real part
of the optical potential in nuclear matter as a function of incident
energy.  The Skyrme interaction has a wrong asymptotic behaviour as a 
function of energy. (This is amplified in \cite{Prakash}.)  Since we
want to devise a momentum dependence which should hold for beam energy 
as high as 1 GeV/nucleon (this would allow investigation of symmetry 
energy at higher than normal nuclear density) we discard the Skyrme 
interaction.

For the purpose of this work we will define nuclear matter to be
an infinite system but without the restriction $N=Z$.
Using the Gogny interaction, we
deduce $U(\rho,\delta,p,\tau)$, the one 
body potential a particle of momentum $p$ and isospin $\tau$ feels
in cold nuclear matter with density $\rho$ and asymmetry $\delta
\equiv\frac{\rho_n-\rho_p}{\rho_n+\rho_p}$.  One then
generalises to $U$ in the case of heavy ion collisions.  For
BUU calculations $U$ is the only quantity needed.  But it is useful
to also have an expression for $V(\rho,\delta)$, the potential energy 
density in cold nuclear matter with a given density 
$\rho$ and asymmetry $\delta$.  This 
allows one to deduce $E/A$ as a function of $\rho$ and $\delta$
which is, of course, of importance.  The expression for $V(\rho,\delta)$
can also be generalised to the case of heavy ion collisions and can be
used to check, for example, the accuracy of energy conservation in
a BUU simulation.

\vskip 0.2in
\epsfxsize=3.5in
\epsfysize=5.0in
\centerline{\rotatebox{270}{\epsffile{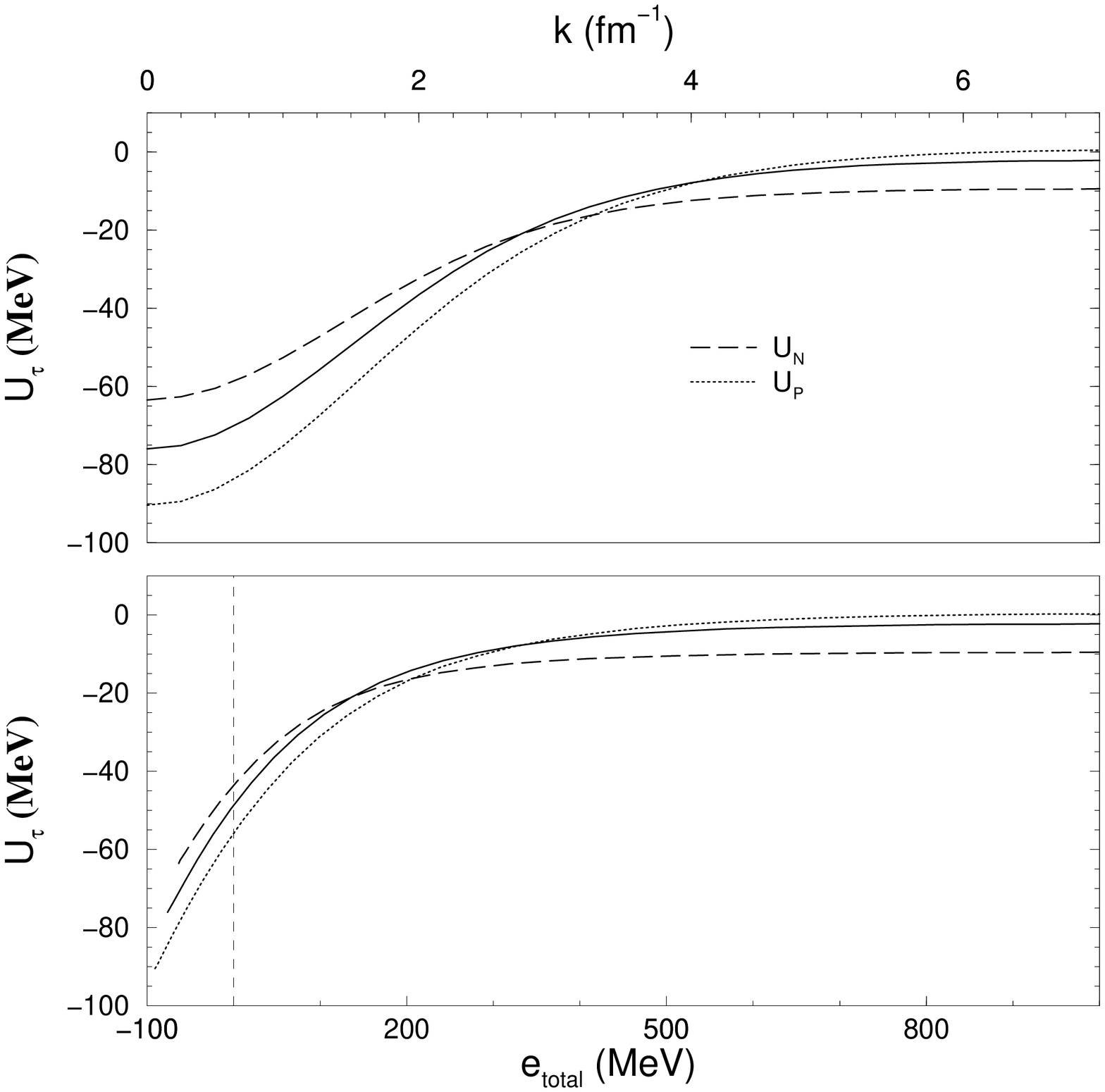}}}
\vskip 0.4in

Fig.1: {\it Single particle potential $U$ with respect to
$k$ and total single particle energy ($e$) for neutron and 
proton for $\delta$=0.4. The solid line is the single particle 
potential for symmetric matter; here $\rho = 0.16 fm^{-3}$.}
\vskip 0.25in

Normally nuclear matter denotes an infinite nucleus with $N=Z$.
Total potential energy in cold matter is deduced from 
\begin{eqnarray}
V_T &=& \frac{1}{2}\sum_{p_1\sigma_1\tau_1p_2\sigma_2\tau_2}
\langle \vec p_1,\sigma_1,\tau_1,\vec p_2,\sigma_2,\tau_2|v(r) \\ \nonumber
 && \left(|\vec p_1,\sigma_1,\tau_1,\vec p_2,\sigma_2,\tau_2\rangle - \right.
\left.|\vec p_2,\sigma_2,\tau_2,\vec p_1,\sigma_1,\tau_1\rangle\right)
\end{eqnarray}
where,
\begin{eqnarray}
v(r) &=& \sum_{i=1,2}(W+BP_{\sigma}-HP_{\tau}-MP_{\sigma}P_{\tau})_i
e^{-r^2/\mu_i^2} \\ \nonumber
&& +t_0(1+P_{\sigma})\rho^{\alpha}(\frac{\vec r_1+
\vec r_2}{2})\delta(\vec r_1-\vec r_2)
\end{eqnarray}
There are two finite range
Gaussians and a density dependent zero-range force.
The values of the parameters
are given in \cite{Gogny}.  The one body potential $U(\rho,\delta,p,\tau)$
is obtained from 

$U(\rho,\delta,p,\tau)=\sum_{p'\sigma'\tau'}   
\langle \vec p,\sigma,\tau,\vec p~',\sigma',\tau'|v(r)
(|\vec p,\sigma,\tau,\vec p~',\sigma',\tau'\rangle
-|\vec p~',\sigma',\tau',\vec p,\sigma,\tau\rangle)$ 
plus rearrangement
term which for nuclear matter is $(3/2)t_0\alpha\rho^{\alpha-1}(1/4)\rho^2
(1-\delta^2)$.

The momentum dependence in $U$ comes entirely from the exchange term
of the finite range part,
i.e., from $\langle\vec p,\vec p~'|e^{-r^2/\mu^2}|\vec p~',\vec p\rangle$.  
Except for the
momentum dependent part, very simple expressions for $U$ and $V_T/A$,
the potential energy per particle are obtained for the Gogny potential.
Thus $U(\rho,\delta,p,\tau)=X+Y+Z$ where $X$ arises from the direct
term of the finite range interaction, $Y$ arises from the $t_0$ term
(density dependent two body term) and $Z$ from the exchange term
of the finite range interactions.  For a given $\tau$, these are:
\begin{eqnarray}
X &=& \rho \left(\sum_{i=1,2}\pi^{3/2}\mu_i^3(W+B/2)_i\right)
-\rho_{\tau} \left(\sum_{i=1,2} \pi^{3/2}\mu_i^3(H+M/2)_i\right) \\
Y &=& \frac{3}{2} t_0\rho^{\alpha}(\rho-\rho_{\tau})+ 
\frac{3}{2} t_0\alpha\rho^{\alpha-1} \frac{1}{4}\rho^2(1-\delta^2) \\
{\rm and} \  Z &=& \sqrt{\pi}\left[\sum_{i=1,2}Z_i(p,\tau)(-W-2B+H+2M)_i\right.
\left.+\sum_iZ_i(p,\tau')(H+2M)_i\right].
\end{eqnarray}
where,
\begin{eqnarray}
Z_i(p,\tau) &=& \frac{1}{\mu_ik} \left[e^{-(\mu_i(k_F(\tau)-k)/2)^2}\right.
\left.-e^{(-\mu_i(k_F(\tau)-k)/2)^2}\right] \\ \nonumber
&& + \frac{\sqrt{\pi}}{2} \left[erf(\frac{\mu_i}{2}(k_F(\tau)-k) \right.
\left.+erf(\frac{\mu_i}{2}(k_F(\tau)+k)\right]. \nonumber
\end{eqnarray}
Here $\tau'\neq \tau$, the isospin of the particle whose
one body potential is being sought.

Similarly, $V_T/A$=potential energy per particle, has contributions 
from the direct term of the finite range force, from the density
dependent $t_0$ term and the exchange term of the finite range
force.  Denoting the first two by $X'$ and $Y'$ respectively, explicit 
expressions for these are:
\begin{eqnarray}
X' &=& \rho[\sum_{i=1,2}\pi^{3/2}\mu_i^3(\frac{W}{2}
+\frac{B}{4}-\frac{H}{4} -\frac{M}{8})_i] 
-\rho\delta^2[\sum_{i=1,2}(\frac{H}{4}+\frac{M}{8})_i] \\
Y' &=& \frac{3}{8}t_0\rho^{\alpha+1}(1-\delta^2).
\end{eqnarray}

\vskip 0.2in
\epsfxsize=3.5in
\epsfysize=4.0in
\centerline{\rotatebox{270}{\epsffile{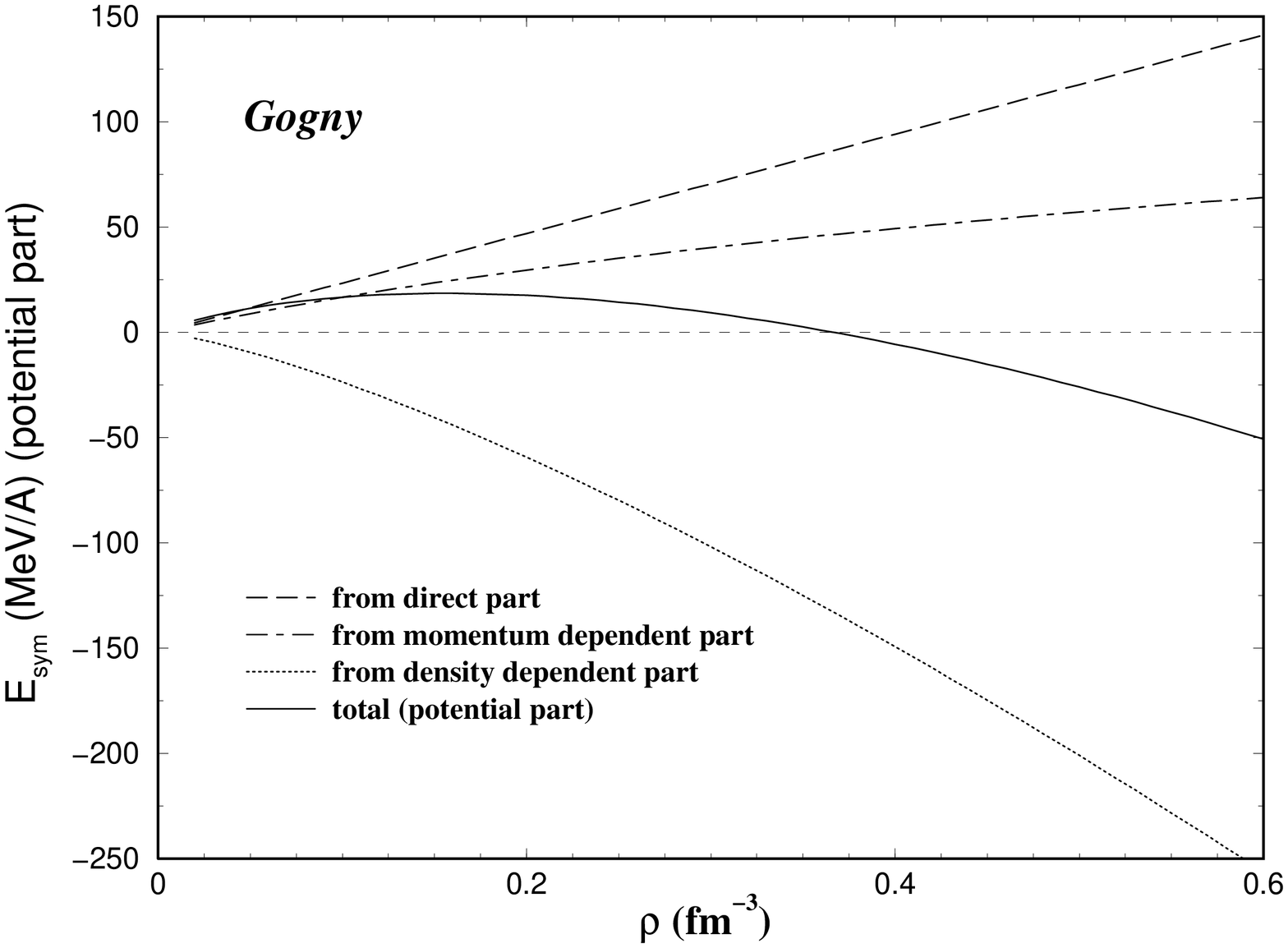}}}
\vskip 0.4in

Fig.2 {\it $E_{sym}(\rho)$ as a function of $\rho$.}
\vskip 0.2in

We do not write down the explicit expression for $Z'$.  It is obtained
from the expression of $Z$ above after a further integration over $p$,
sum over $\sigma, \tau$ and dividing the answer by 2.

The one body potentials as a function of $k$ for neutrons and protons
in cold matter as predicted by the Gogny potential for $\delta=0$
and $\delta=0.4$ are shown in Fig. 1.  They are quite similar
to other calculations of $U(\rho,\delta,p,\tau)$ available
in the literature.  We compare, in particular, to the $U(\rho,\delta,p,\tau)$
given in Fig.4 in \cite{Bombaci2}.  That figure is for $\rho=.17$
and obtained from Brueckner-Hartee-Fock calculation.  With Gogny interaction
we find that the equation of state ({\rm EOS}) of asymmetric nuclear matter can be written as
$\frac{E}{A}(\rho,\delta)\approx \frac{E}{A}(\rho,0)+E_{sym}(\rho)\delta^2$,
in agreement with the empirical parabolic law found by all many body theories.  
Different calculations depending on the many body approaches and the interactions used, 
give very different behaviors for $E_{sym}(\rho)$, especially at high densities. 
In some, such as, the Relativistic Mean Field model\cite{rmf} and the 
Brueckner-Hartree-Fock\cite{Bombaci2}, it is a continuously rising function of $\rho$, 
In others, such as, the Variational Many-Body Approach\cite{vma},
it rises in the beginning and then begins to fall. Within the Hartree-Fock approach using 
all 86 Skyrme effective interactions widely used currently in nuclear structure studies, 
it was found that about ${1/3} (2/3)$ of them lead to symmetry energies in 
the first (second) category\cite{stone}. Gogny interaction with the default 
parameters has behaviour in the second category. It is interesting to see what 
causes the fall and how the parameters of the interaction can be
changed minimally to alter this behaviour.  In Fig.2
we plot contributions to $E_{sym}(\rho)$ from the Gogny interaction and
we show separately the term coming from (a) the direct term of the
finite range part, (b) the exchange term of the finite range part and
(c) the density dependent zero range part.  It is the latter
that causes the bending over.  For example, if we choose the density
dependent term to be $\delta$ independent, $E_{sym}(\rho)$ will continue
to rise.

Lastly, we write the potential energy density due to Gogny interaction
in a form that is common practice in BUU literature. Thus

\begin{eqnarray}
V(\rho,\delta) &=& -52.41(\frac{\rho^2}{\rho_0}) +37.70(\frac{\rho^2}{\rho_0})
\delta^2+\frac{102.6}{\sigma+1}\frac{\rho^{\sigma+1}}{\rho_0^{\sigma}}(1-
\delta^2) \nonumber \\
&& + \frac{1}{\rho_0}\sum_{\tau,\tau'}\int\int f_{\tau}(\vec r,\vec p)
f_{\tau'}(\vec r,\vec p')\left[(-41.51+46.02\delta_{\tau,\tau'}) \right.
\left. e^{-(\vec k-\vec k')^2\mu_1^2/4} \right. \nonumber \\
&& \left. +(-38.62+17.25\delta_{\tau\tau'}) \right.
\left.e^{-(\vec k-\vec k')^2\mu_2^2/4}\right]d^3pd^3p'.
\end{eqnarray}
Here, as in previous work, all quoted numbers are in MeV, $\rho_0=0.16
fm^{-3}$.  Also $\sigma=\alpha$(Gogny)+1=4/3.

Given that the momentum dependence generated by the Gogny potential
comes from two Gaussians, we proceed to find a simpler version for
momentum dependent potential for asymmetric matter.  We will extend
the parametrisation of \cite{Welke,Gale2}. Subsequently we will
find an even simpler version, the kind that was used in first
applications of momentum dependent potentials for heavy ion collisions.

\section{A simple momentum dependent potential for BUU calculations}
The simplest generalisation of the potential energy density of Eq.(5.4)
of \cite{Welke} to asymmetric nuclear matter is
\begin{eqnarray}
V(\rho,\delta) &=& \frac{A_1}{2\rho_0}\rho^2+
\frac{A_2}{2\rho_0}\rho^2\delta^2+
\frac{B}{\sigma+1}\frac{\rho^{\sigma+1}}{\rho_0^{\sigma}}(1-x\delta^2)
 \nonumber \\
&& +\frac{1}{\rho_0}\sum_{\tau,\tau'}C_{\tau,\tau'}
\int\int d^3pd^3p'\frac{f_{\tau}(\vec r,
\vec p)f_{\tau'}(\vec r,\vec p')}{1+(\vec p-\vec p')^2/\Lambda^2}
\end{eqnarray}
The parameter $x$ is introduced to cover the largely uncertain behavior of 
nuclear symmetry energy $E_{sym}(\rho)$ as discussed in the previous section.  
For the choice of $x=1$ (same as in Gogny) in the term containig $B$,
the symmetry energy will bend over beyond a density $\rho$; for the 
choice $x=0$ the symmetry energy will continue to rise with density.
In the above, $C_{1/2,1/2}=C_{-1/2,-1/2}=C_{like}$ 
and $C_{1/2,-1/2}=C_{-1/2,1/2}=C_{unlike}$.
In terms of interactions between like and unlike particles, the above
equation is equivalent to
\begin{eqnarray}
V(\rho,\delta) &=& \frac{A_u\rho_n\rho_p}{\rho_0}+
\frac{A_l}{2\rho_0}(\rho_n^2+\rho_p^2)+
\frac{B}{\sigma+1}\frac{\rho^{\sigma+1}}{\rho_0^{\sigma}}(1-x\delta^2)
 \nonumber \\
&& +\frac{1}{\rho_0}\sum_{\tau,\tau'}C_{\tau,\tau'}
\int\int d^3pd^3p'\frac{f_{\tau}(\vec r,
\vec p)f_{\tau'}(\vec r,\vec p')}{1+(\vec p-\vec p')^2/\Lambda^2}
\end{eqnarray}
where $A_1=(A_u+A_l)/2$ and $A_2=(A_l-A_u)/2$. The one-body potential
needed for BUU computations is given by
\begin{eqnarray}
U(\rho,\delta,\vec p,\tau) &=& A_u\frac{\rho_{\tau'}}{\rho_0}
+A_l\frac{\rho_{\tau}}{\rho_0}+
B(\frac{\rho}{\rho_0})^{\sigma}(1-x\delta^2)-x\frac{B}{\sigma+1}\frac{\rho^{
\sigma+1}}{\rho_0^{\sigma}}\frac{d\delta^2}{d\rho_{\tau}} \nonumber \\
&& +\frac{2C_{\tau,\tau}}{\rho_0}
\int d^3p'\frac{f_{\tau}(\vec r,\vec p')}{1+(\vec p-\vec p')^2/\Lambda^2}
+\frac{2C_{\tau,\tau'}}{\rho_0}
\int d^3p'\frac{f_{\tau'}(\vec r,\vec p')}{1+(\vec p-\vec p')^2/\Lambda^2}
\end{eqnarray}
In the above $\tau\neq\tau'$ and $\frac{\partial\delta^2}{\partial\rho_n}=
\frac{4\delta\rho_p}{\rho^2}$ and $\frac{\partial\delta^2}{\partial\rho_p}=
-\frac{4\delta\rho_n}{\rho^2}$.

The constants appearing in Eqs.(3.1) and (3.2) will be fixed by ensuring
that properties of cold nuclear matter are reproduced.  There
$f_{\tau}(\vec r,\vec p)=\frac{2}{h^3}\Theta(p_f(\tau)-p)$.  The integration
in Eq.(3.1) is facilitated by noting that for a fixed $\vec p\equiv(\vec p_1-
\vec p_2)/2$, the centre of mass momentum can be integrated out to give
\begin{eqnarray}
\int_0^{p_f(\tau)}\int_0^{p_f(\tau')}d^3p_1d^3p_2g(\vec p) &=&
\int_0^{q_f} \left[ \frac{16\pi}{3} \right.
\left. (p_f^3(\tau)+p_f^3(\tau'))-8\pi p(p_f^2(\tau)+p_f^2(\tau')) \right. \nonumber \\
&&\left. +\frac{16\pi}{3}p^3-\frac{\pi}{p}(p_f^2(\tau)-p_f^2(\tau'))^2 \right] g(\vec p)d^3
p
\end{eqnarray}
where $q_f = (p_f(\tau)+p_f(\tau'))/2 $.

\vskip 0.2in
\epsfxsize=3.5in
\epsfysize=5.0in
\centerline{\rotatebox{270}{\epsffile{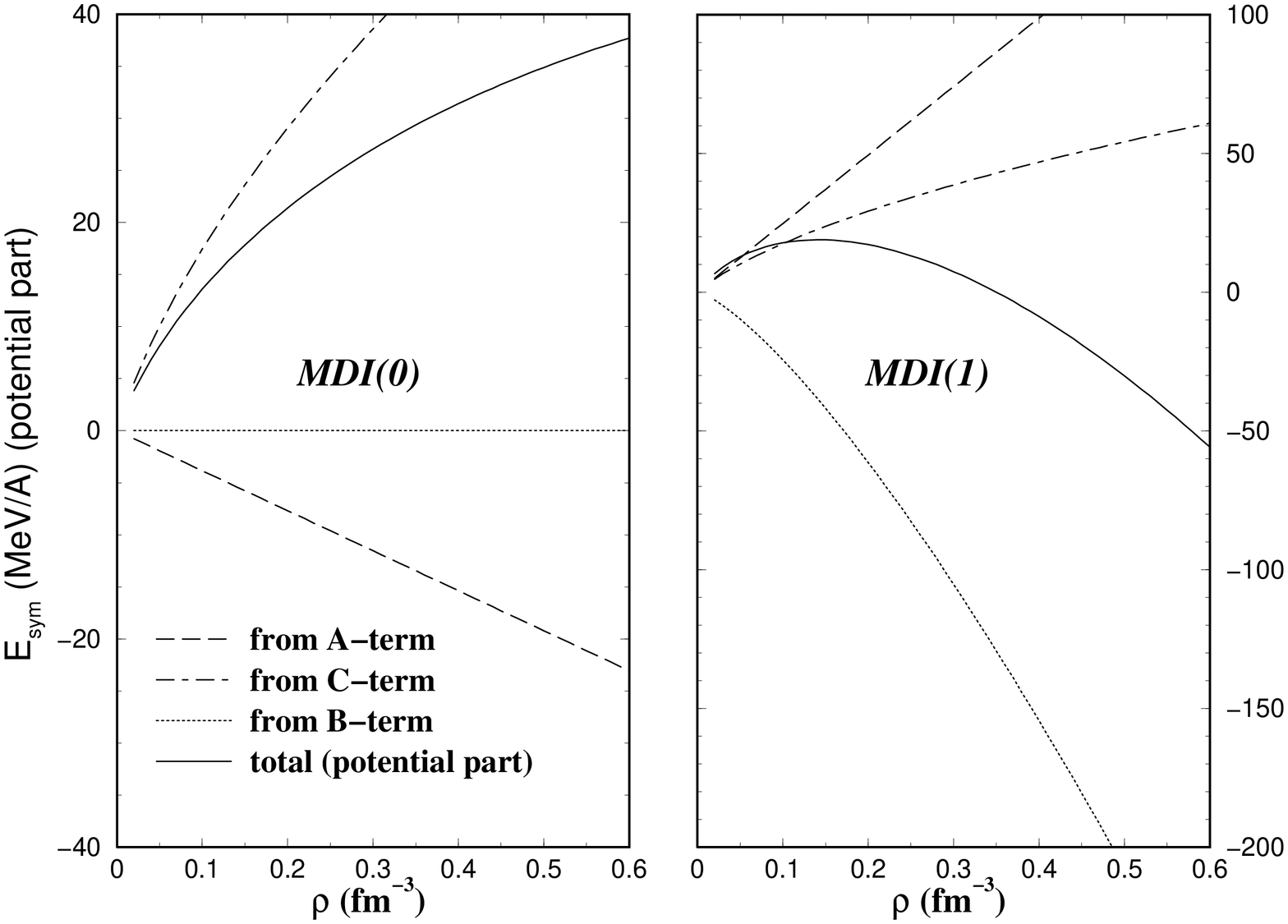}}}
\vskip 0.4in

Fig.3 {\it $E_{sym}(\rho)$ as a function of $\rho$ for the 
two choices of
$x$; 0(left panel) and 1(right panel).}
\vskip 0.25in

For completeness, for cold matter, we write down the values of
the integrals appearing in Eqs.(3.1) and (3.2)
\begin{eqnarray}
\int\int d^3pd^3p'\frac{f_{\tau}(\vec r, \vec p)f_{\tau'}(\vec r,\vec p')}
{1+(\vec p-\vec p')^2/\Lambda^2} &=& \left(\frac{2}{h^3}\right)^2 ~\frac{4}{3} ~\pi^2 ~\Lambda^2
\left[\{q_f-\frac{\Lambda}{2} {\rm arctan}(\frac{2q_f}{\Lambda})\}4(p_f^3(\tau)+p_f^3(\tau')) \right. \nonumber \\
&& \left. - \{3(p_f^2(\tau)+p_f^2(\tau'))+\frac{\Lambda^2}{2}\} {q_f}^2 + {q_f}^4 \right. \nonumber \\
&& \left. + \{\frac{3\Lambda^2}{4}(p_f^2(\tau)+p_f^2(\tau'))+\frac{\Lambda^4}{8} \right.
\left. -\frac{3}{8}(p_f^2(\tau)-p_f^2(\tau'))^2\} ln (1+\frac{4{q_f}^2}{\Lambda^2})\right]
\end{eqnarray}
Similarly, the value of the integral in Eq.(3.3) is
\begin{eqnarray}
\int d^3p'\frac{f_{\tau}(\vec r,\vec p')}{1+(\vec p-\vec p')^2/\Lambda^2} &=& \frac{2}{h^3}\pi \Lambda^3
\left[ \frac{p_f^2(\tau)+\Lambda^2-p^2}{2p\Lambda} ln \frac{(p+p_f(\tau))^2+\Lambda^2}{(p-p_f(\tau))^2+\Lambda^2} \right. \nonumber \\
&& \left. + \frac{2p_f(\tau)}{\Lambda} - 2\{{\rm arctan}\frac{p+p_f(\tau)}{\Lambda}-{\rm arctan}\frac{p-p_f(\tau)}{\Lambda}\} \right]
\end{eqnarray}

\vskip 0.2in
\epsfxsize=3.5in
\epsfysize=5.0in
\centerline{\rotatebox{270}{\epsffile{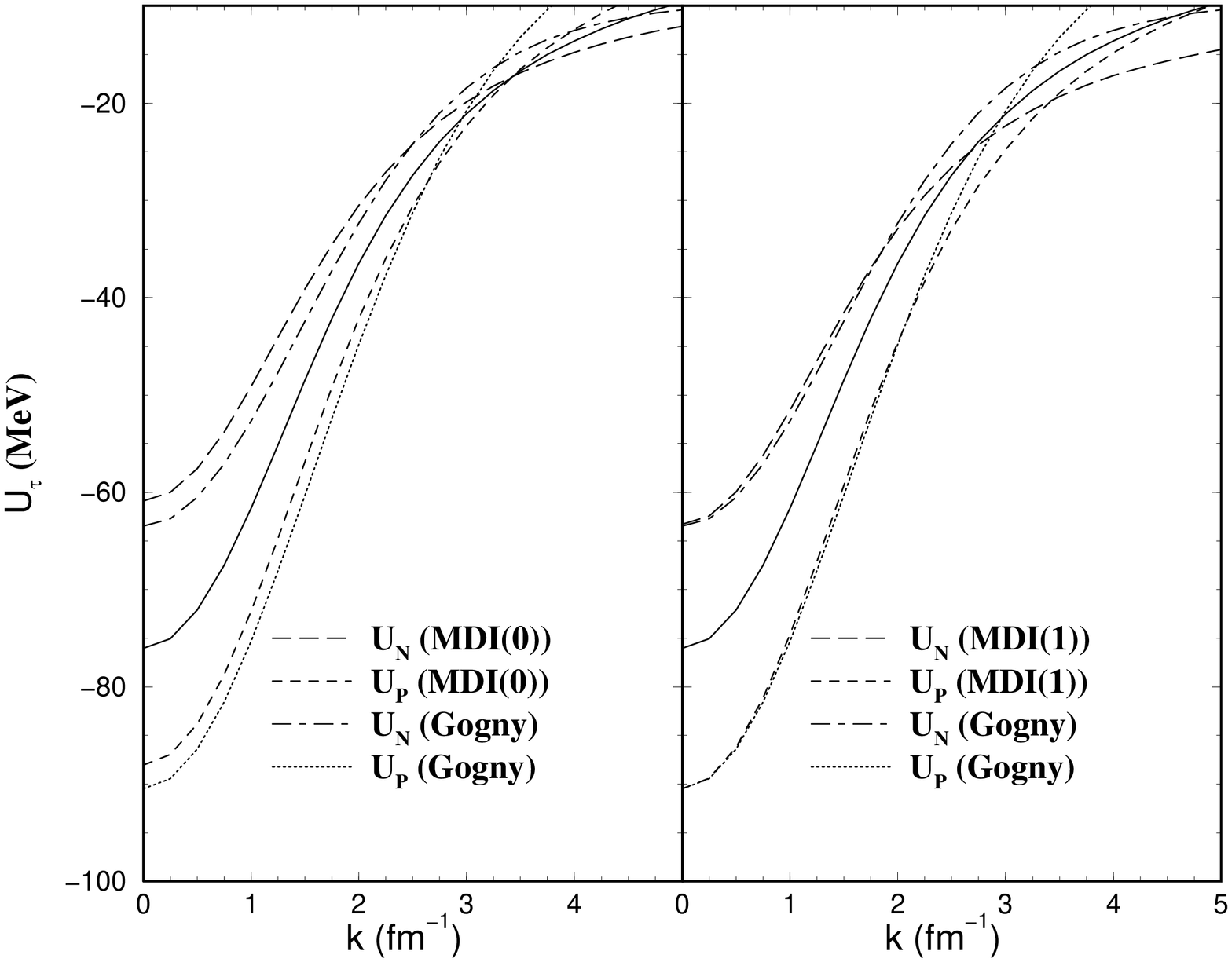}}}
\vskip 0.4in

Fig.4 {\it One body potential in cold matter for $\delta=0.4$. 
As in fig. 3,
here also the two panels are for $x$=0 and 1. Comparison with one-body
potential as obtained with a Gogny interaction has also been shown.
The solid line is the single particle potential for symmetric matter
with MDI.}
\vskip 0.25in

We fix the force parameters by first optimally reproducing the variation
of $U(\rho_0,\delta,p)$ with $p$ with that obtained by using a Gogny force
(Brueckner-Hartee-Fock calculation gives similar results
\cite{Bombaci2}).  This fixes $\Lambda$ of Eq.(3.3).  Other 
parameters are then fixed by values of saturation density (0.16 $fm^{-3}$),
binding energy (16 MeV) at saturation density, compressibility ($\approx$
210 MeV) of N=Z nuclear matter at saturation density and symmetry energy
at $N=Z$ ($\approx$ 30 MeV). The values of the parameters for two choices 
(1) $x$=0 (denoted by MDI(0)) and (2) $x$=1 (denoted by MDI(1)) in Eq.(3.2) 
as in Gogny (this causes the value of the symmetry energy to bend over 
as a function of density) and are given in the table.  Fig. 3 shows the 
behaviour of symmetry energy for the two choices. In fig. 4 we show the 
one body potential in cold matter for $\delta=0.4$.

Before we close this section we like to add that BUU calculations with
momentum dependent interactions are only slightly more complicated
in the asymmetric case compared to the case where no distinction is
made between neutrons and protons.  In numerical solutions of Boltzmann
equations, phase space densities are simulated by test particles,
characterised by a position and a momentum.  Now there will also  
be a tag on their charge but no major addition to the codes
\cite{Zhang} are needed.

\section{Reduction to the GBD form}

Although we do recommend the full formalism of the above section be
implemented, it is possible to reduce the above to a GBD (Gale, Bertsch
and Das Gupta) form.  A GBD potential, extended to asymmetric matter,
already exists and is called BGBD \cite {Bombaci}.  There the extensions
were made such that in the $N=Z$ case one gets back exactly the original
parameters \cite{Gale1}.  The parameter $x$ (eq.(3.1) was chosen to be 
1/15.  In contrast, here we have chosen the momentum dependence
of the Gogny potential as a reference curve and chosen $x$ at 0 or 1.
The values of the paramaters of GBD(0) ($x$=0), GBD(1) ($x$=1) and BGBD 
($x$=1/15) are given in the table.

We write the potential energy density coming
from the momentum dependent part as
\begin{eqnarray}
V_{mom}(\rho,\delta)=\frac{1}{\rho_0}\sum_{\tau,\tau'}C_{\tau,\tau'}\rho_{\tau}
\int\frac{f_{\tau'}(\vec r,\vec p)d^3p}{1+(\vec p-<\vec p>_{\tau})^2/\Lambda^2}
\end{eqnarray}
The one-body potential generated by this piece of potential energy density 
for given $\rho, \delta, \vec p$ and a given $\tau$ is
\begin{eqnarray}
U_{mom}(\rho,\delta,\vec p,\tau) &=& \frac{C_{\tau,\tau}}{\rho_0}
\left[\int \frac{f_{\tau}(\vec r,\vec p')d^3p'}{1+(\vec p'-<\vec p>_{\tau})^2/\Lambda^2} \right.
\left.+\frac{\rho_{\tau}}{1+(\vec p-<\vec p>_{\tau})^2/\Lambda^2}\right]
\nonumber \\
&&+\frac{C_{\tau,\tau'}}{\rho_0}
\left[\int\frac{f_{\tau'}(\vec r,\vec p')d^3p'}{1+(\vec p'-<\vec p>_{\tau'})^2/\Lambda^2} \right.
\left.+\frac{\rho_{\tau'}}{1+(\vec p-<\vec p>_{\tau'})^2/\Lambda^2}\right]
\end{eqnarray}
In the above equation, $\tau'\neq\tau$.
As expected, the values of the constants in the force will have to
be recalculated to reproduce the saturation properties.  These are given 
in the table.  The GBD potential $U(\rho,\delta,p,\tau)$ is plotted in Fig.5.
This does not track the Gogny potential as faithfully as the more 
sophisticated version of section III does.
The reader might wonder why the value of $\Lambda$ is 
significantly bigger in GBD as opposed to in MDI(0) and MDI(1).  
We have tried to fit the variation of $U$ with $k$ as obtained in
Gogny potential (or the Brueckner-Hartree-Fock calculation) by adjusting
the value of $\Lambda$.  For $p$ close $p_F$, the
contribution to $U(p)$ in section III comes mainly from $p'$ near 
$p_F$ whereas in GBD what counts is $\mid \vec p-<\vec p>_{ave} \mid$
: since $,\vec <p>_{ave}$ is zero, one requires a 
different value of $\Lambda$
to mimic the variation with $p$.  This point was not appreciated in
\cite {Welke}.

\vskip 0.2in
\epsfxsize=3.5in
\epsfysize=5.0in
\centerline{\rotatebox{270}{\epsffile{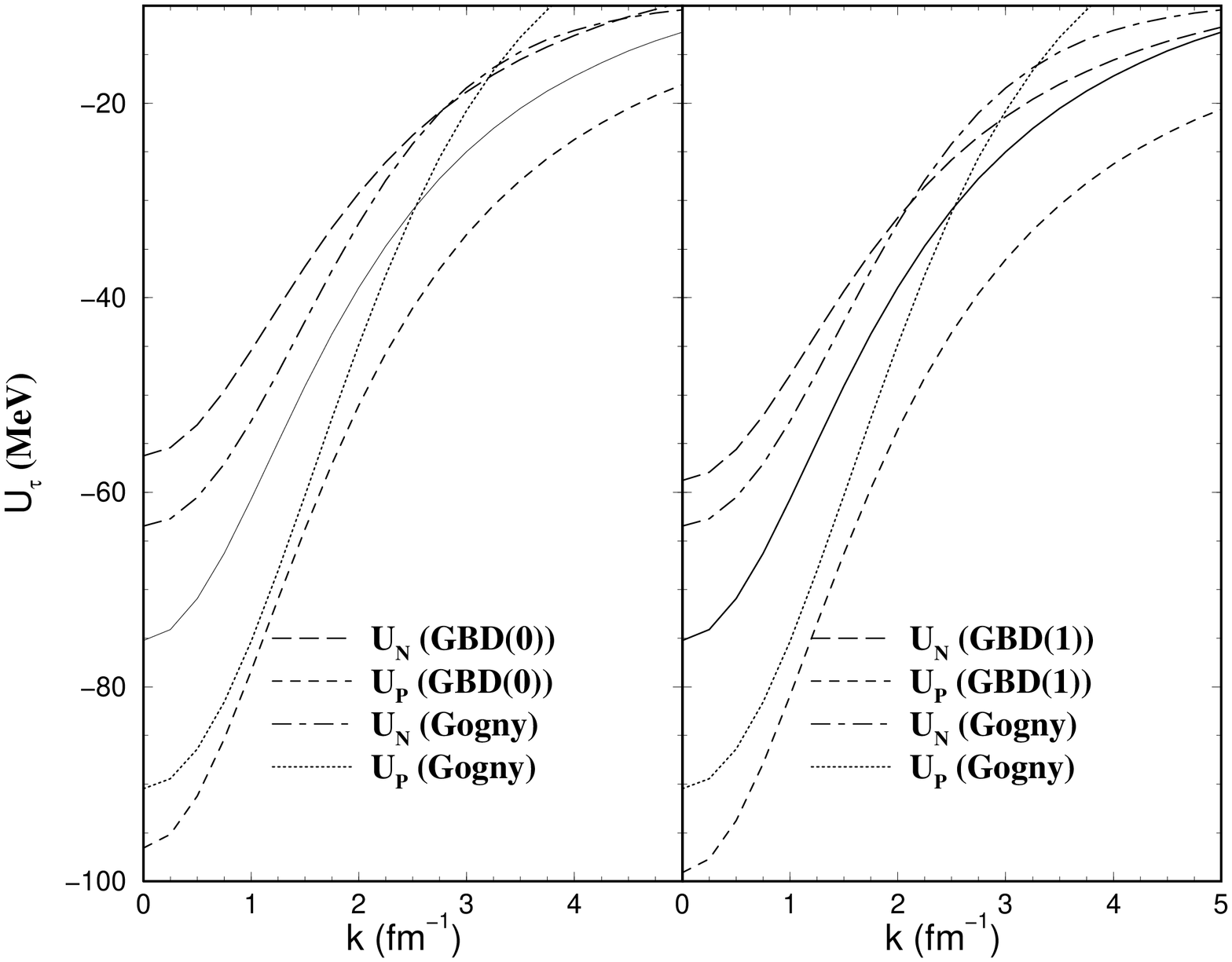}}}
\vskip 0.4in

Fig.5 {\it One body potential in cold matter for $\delta=0.4$. 
As in fig. 4,
here also the two panels are for $x$=0 and 1. Comparison with one-body
potential as obtained with a Gogny interaction has also been shown.
The solid line represents the single particle potential
for symmetric matter with the GBD potential.}
\vskip 0.25in

\section{Momentum dependence of the symmetry potential}

The single particle potentials derived in the previous
sections can be used 
directly in transport model calculations. They combine
the density, momentum and isospin
dependences of both the isoscalar and symmetry
potentials in a nontrivial way.
In this section we evaluate the strength
of the momentum dependence of the symmetry potential. 
To the leading order in $\delta$,
the single nucleon potential can be cast to the form
\begin{equation}
U_{n/p}(\rho,\vec{p},\delta)\approx U_0(\rho,\vec{p})
\pm U_{sym}(\rho,\vec{p})\delta
\end{equation}
in accordance with the Lane potential\cite{lane},
where the $\pm$ sign is for neutrons and
protons, respectively. Thus the symmetry potential can
be evaluated from
$U_{sym}(\rho,\vec{p})=(U_n-U_p)/2\delta$. Shown in
Fig. 6 are the symmetry potentials as
a function of $k$ for the three densities. It is seen
that the $U_{sym}(\rho,\vec{p})$ is 
strongly momentum dependent for $k\leq 5 fm^{-1}$ in all
models considered. Moreover, 
this dependence is particularly stronger at high
densities. By construction, the results 
for Gogny and MDI(1) are very close. By comparing the
results with $x=0$ and $x=1$, it is 
seen that the momentum dependence is rather different
mainly for $\rho/\rho_0=2$. 
This is because the symmetry energies with $x=0$ and
$x=1$ are significantly different only
in the region of $\rho\geq \rho_0$ as shown in Fig. 3.

\vskip 0.2in
\epsfxsize=3.5in
\epsfysize=5.0in
\centerline{\rotatebox{270}{\epsffile{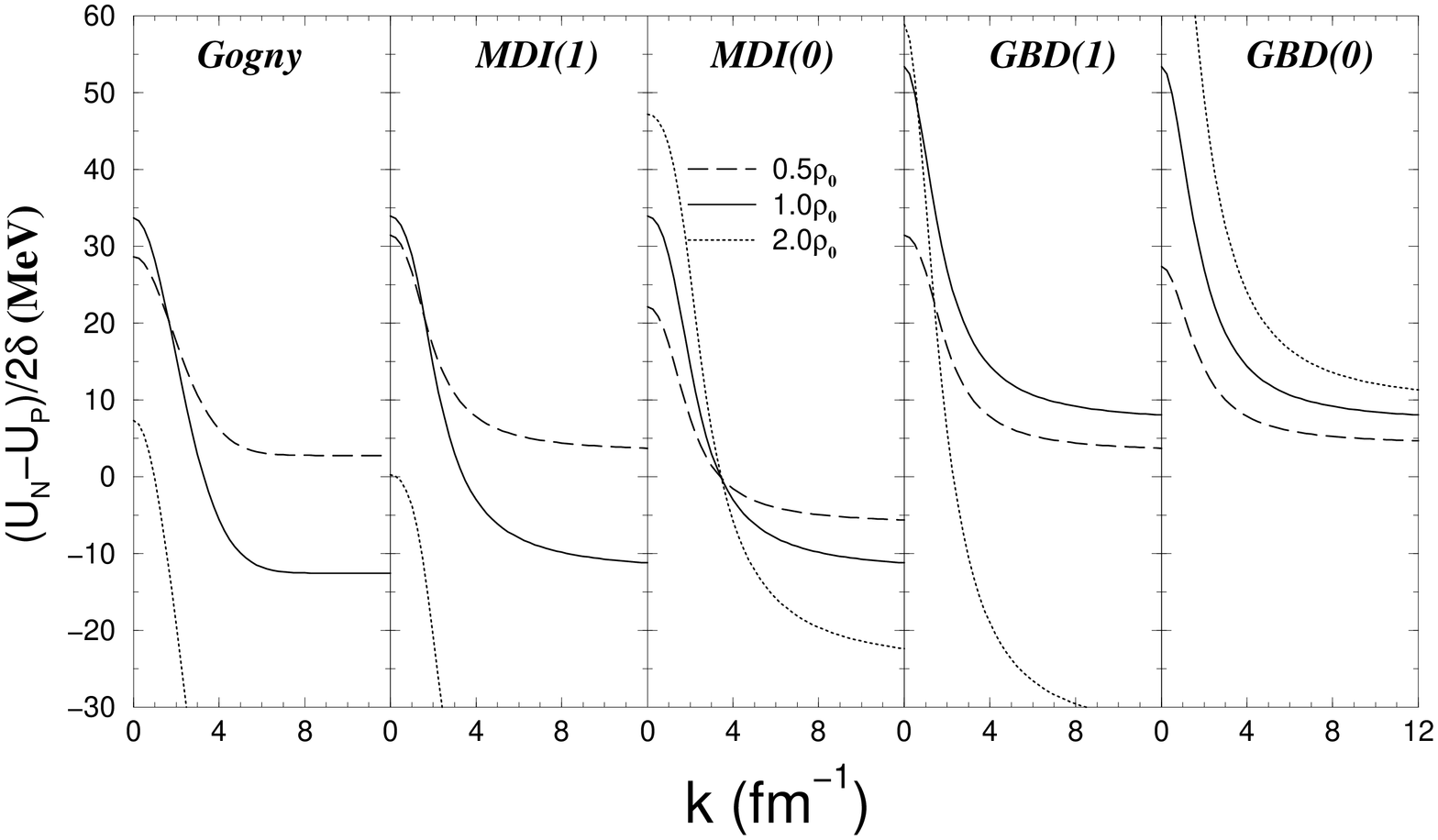}}}
\vskip 0.4in

Fig.6 {\it Dependence of $(U_N - U_P)/2\delta$ on momentum for
different potentials at different densities. All the figures
are for $\delta$=0.4} 
\vskip 0.2in

To our best knowledge, the symmetry potential has been
assumed to be momentum independent
in all previous studies. Our results above indicate
that this is only a good approximation at 
high momenta. At momenta less than about 1 GeV/c, the
momentum dependence of the symmetry
potential is important.

\section{Summary}
The goal of this work was to obtain parameters of a momentum dependent
mean field potential which is applicable to highly isospin asymmetric nuclear 
matter but easy enough to use in a transport model calculation for heavy ion 
collisions at intermediate to high energies. We used the Gogny interaction as
a guide.  Published results of Brueckner-Hartree-Fock calculations
were also used for choosing parameters.  We have two versions.
They both are quite flexible in the sense that parameters
can be easily chosen according to that experimental data
on binding energy, compressibility etc.
In each of these 
versions we have proposed two sets of parameters: one, in which the 
symmetry energy continues to rise as a function of density and another one where the
symmetry energy first rises and then falls off with density.  
Our opinion is that introducing momentum dependence in symmetry potential
will not not make transport model simulations much harder or longer
than they already are. Implementations of these potentials in BUU transport models are
in progress. 

Transport models with the momentum dependent symmetry potentials are more reliable tools for 
investigating the density dependence of nuclear symmetry energy and thus 
the EOS of neutron-rich matter. Knowledge on the symmetry energy $E_{sym}(\rho)$ is essential 
for understanding not only the structure of radioactive nuclei 
but also many key issues in astrophysics. For instance, the $E_{sym}(\rho)$ determines 
uniquely the proton fraction in neutron stars at $\beta$ equilibrium\cite{Bombaci}. 
A continuously rising symmetry energy leads to a growing proton fraction with increasing density, 
thus allowing for the fast cooling of protoneutron stars through 
the direct URCA process\cite{lat91}. A falling 
symmetry energy at high densities forbides the direct URCA process to happen; moreover, it 
favors the formation of pure neutron domains in the cores of 
neutron stars\cite{kut94}. Nuclear reactions induced by neutron-rich nuclei provide a great 
opportunity to pin down the density dependence of nuclear symmetry energy. Among the experimental 
observables that have been found to be sensitive to the symmetry potential, the neutron/proton
ratio of pre-equilibrium nucleon emissions, neutron-proton differential flow and correlation 
functions, as well as the proton elliptic flow at high transverse momenta are expected to be
most sensitive to the momentum dependence of the symmetry potential. These observables will
be studied with the improved BUU transport models using the momentum dependent symmetry potentials.
These results will be reported in a forthcoming publication. 

\section{Acknowledgement}
S. Das Gupta acknowledges a very useful discussion with J.M. Pearson. 
Bao-An Li acknowledges the warm hospitality he received at McGill University 
where this work started. This research is supported in part by the Natural Sciences and 
Engineering Research Council of Canada and by the National Science Foundation of the 
United States under grant No. PHY-0088934.

\begin{table}
\caption{The values of the parameters of different interactions.
Also gives the values $K$ in the respective 
parametrisations. The saturation density for all parametrisation
is $0.16 fm^{-3}$, except for BGBD(1/15) it is $0.163 fm^{-3}$. 
The binding energy and the total symmetry energy including 
contribution from kinetic part are -16 MeV/A and
31.623 MeV/A respectively, in each case.}
\vspace {0.5in}
\begin{tabular}{cccccc}
\hline
\multicolumn{1}{c}{Parameters} &
\multicolumn{1}{c}{MDI(0)} &
\multicolumn{1}{c}{MDI(1)} &
\multicolumn{1}{c}{GBD(0)} &
\multicolumn{1}{c}{GBD(1)} &
\multicolumn{1}{c}{BGBD(1/15)} \\
\hline
$A_u$&-95.98&-187.27&-109.85&-299.69&-192.0 \\
$A_l$&-120.57&-29.28&-191.30&-1.46&-96.0 \\
$B$&106.35&106.35&205.66&205.66&203.3 \\
$C_u$&-103.40&-103.40&-118.80&-118.80&-84.53 \\
$C_l$&-11.70&-11.70&-26.26&-26.26&-65.472 \\
$\sigma$&4/3&4/3&7/6&7/6&7/6 \\
$\Lambda$&$1.0{p_f}^{(0)}$&$1.0{p_f}^{(0)}$&$1.5{p_f}^{(0)}$&$1.5{p_f}
^{(0)}$&$1.5{p_f}^{(0)}$ \\
\hline
$K$&210.68&210.68&214.68&214.68&215.0 \\
\end{tabular}
\end{table}

\end{document}